\documentclass[manuscript]{acmart}

\def\arxiv{1}

\AtBeginDocument{%
  }

\copyrightyear{2025}
\acmYear{2025}
\setcopyright{rightsretained}
\acmConference[SIGCSE TS 2025]{Proceedings of the 56th ACM Technical
Symposium on Computer Science Education V. 1}{February 26-March 1,
2025}{Pittsburgh, PA, USA}
\acmBooktitle{Proceedings of the 56th ACM Technical Symposium on Computer
Science Education V. 1 (SIGCSE TS 2025), February 26-March 1, 2025,
Pittsburgh, PA, USA}
\acmDOI{10.1145/3641554.3701865}
\acmISBN{979-8-4007-0531-1/25/02}

\usepackage{multicol}
\usepackage{xspace}
\usepackage{amsmath}

\usepackage{siunitx}
\usepackage{graphicx}
\usepackage{subcaption}
\usepackage{booktabs}       %

\usepackage{mathtools}

\usepackage{color-edits}
\addauthor[Aditya]{ap}{orange}
\addauthor[Ram]{rd}{blue}
\addauthor[David Lee]{dl}{green}
\addauthor[David Kempe]{dk}{red}

\newcommand{\dlnote}[1]{\dlcomment{#1}}

\usepackage[ruled,commentsnumbered,linesnumbered,noend]{algorithm2e} 
\DontPrintSemicolon
\SetArgSty{textrm} %

\usepackage{bm}

\usepackage{xcolor}
\usepackage{amsthm}

\newtheorem{problem}{Problem}[section]

\usepackage[inline]{enumitem}

\usepackage{algorithm2e}

\usepackage{listings}
\definecolor{codegreen}{rgb}{0,0.6,0}
\definecolor{codegray}{rgb}{0.5,0.5,0.5}
\definecolor{codepurple}{rgb}{0.58,0,0.82}
\definecolor{bg}{rgb}{0.95,0.95,0.92}

\lstdefinestyle{mystyle}{
    backgroundcolor=\color{bg},
    commentstyle=\color{codegreen},
    keywordstyle=\color{magenta},
    numberstyle=\tiny\color{codegray},
    stringstyle=\color{codepurple},
    basicstyle=\ttfamily\footnotesize,
    breakatwhitespace=false,         
    breaklines=true,                 
    captionpos=b,                    
    keepspaces=true,                 
    numbers=left,                    
    numbersep=5pt,                  
    showspaces=false,                
    showstringspaces=false,
    showtabs=false,                  
    tabsize=2,
    identifierstyle=\color{codegreen},
    rulecolor=\color{codegreen},
}
\usepackage{array}
\usepackage{booktabs}
\usepackage{multirow}
\usepackage{adjustbox}
\newcolumntype{X}[2]{%
    >{\adjustbox{angle=#1,lap=\width-(#2)}\bgroup}%
    c%
    <{\egroup}%
}

\usepackage{soul}
\usepackage{censor}

\usepackage{rotating}
\usepackage{color, colortbl}

\providecommand{\SET}[1]{\ensuremath{\{ #1 \}}\xspace}
\providecommand{\Set}[2]{\ensuremath{\SET{#1 \mid #2}}\xspace}
\providecommand{\Kth}[1]{\ensuremath{{#1}^{\rm th}}}
\providecommand{\OPT}{\ensuremath{\text{OPT}}}

\usepackage{wasysym}

\usepackage{amsmath,amsfonts,bm}

\def\eqref#1{equation~\ref{#1}}

\def\1{\bm{1}}

\DeclareMathAlphabet{\mathsfit}{\encodingdefault}{\sfdefault}{m}{sl}
\SetMathAlphabet{\mathsfit}{bold}{\encodingdefault}{\sfdefault}{bx}{n}

\DeclareMathOperator*{\argmax}{arg\,max}
\DeclareMathOperator*{\argmin}{arg\,min}

\definecolor{tableoptions}{rgb}{0.85,0.85,0.85}
\definecolor{tabletotal}{rgb}{0.95,0.95,0.95}

\usepackage{xspace}
\newcommand{\dpvis}{dpvis\xspace}

\begin{document}

\title{\dpvis: A Visual and Interactive Learning Tool for Dynamic Programming}

\author{David H. Lee}
\email{dhlee@usc.edu}
\affiliation{%
 \institution{University of Southern California}
 \city{Los Angeles}
 \state{California}
 \country{USA}
}

\author{Aditya Prasad}
\authornote{Work completed while the authors were at the University of Southern California.}
\email{adityaprasad@uchicago.edu}
\affiliation{%
  \institution{University of Chicago}
  \city{Chicago}
  \state{Illinois}
  \country{USA}
}

\author{Ramiro Deo-Campo Vuong}
\authornotemark[1]
\email{ramdcv@cs.cornell.edu}
\affiliation{%
 \institution{Cornell University}
 \city{Ithaca}
 \state{New York}
 \country{USA}
}

\author{Tianyu Wang}
\email{twang659@usc.edu}
\affiliation{
 \institution{University of Southern California}
 \city{Los Angeles}
 \state{California}
 \country{USA}
}

\author{Eric Han}
\email{ejhan@usc.edu}
\affiliation{
 \institution{University of Southern California}
 \city{Los Angeles}
 \state{California}
 \country{USA}
}

\author{David Kempe}
\email{David.M.Kempe@gmail.com}
\affiliation{%
 \institution{University of Southern California}
 \city{Los Angeles}
 \state{California}
 \country{USA}
}

\begin{abstract} %
Dynamic programming (DP) is a fundamental and powerful algorithmic paradigm taught in most undergraduate (and many graduate) algorithms classes.
DP problems are challenging for many computer science students because they require identifying unique problem structures and a refined understanding of recursion.
In this paper, we present \dpvis, a Python library that helps students understand DP through a frame-by-frame animation of dynamic programs.
\dpvis can easily generate animations of dynamic programs with as little as \emph{two lines of modifications} compared to a standard Python implementation.
For each frame, \dpvis highlight the cells that have been read from and written to during an iteration.
Moreover, \dpvis allows users to test their understanding by prompting them with questions about the next operation performed by the algorithm.

We deployed \dpvis as a learning tool in an undergraduate algorithms class, and report on the results of a survey.
The survey results suggest that \dpvis is especially helpful for visualizing the recursive structure of DP. 
Although some students struggled with the installation of the tool (which has been simplified since the reported deployment), essentially all other students found the tool to be useful for understanding dynamic programs.
dpvis is available at \href{https://github.com/itsdawei/dpvis}{https://github.com/itsdawei/dpvis}.
\end{abstract}



\maketitle

\section{Introduction}
Dynamic programming (DP) is an algorithmic technique taught in many introductory algorithms courses as a fundamental part of undergraduate computer science curricula \citep{acm2013computer, kleinberg2005algorithm, cormen:leiserson:rivest:stein}.
DP exploits the recursive structure of combinatorial problems to break them into subproblems.
Then, DP reduces redundant calculations by storing the solutions to prior subproblems in a \emph{subproblem array} and referencing them to solve later subproblems.
As a result, DP often yields polynomial-time algorithms for difficult-looking problems.
(For a more in-depth refresher on DP, see \autoref{sec:background}.)

Anecdotal experience of most algorithms instructors and past research have pointed to pervasive difficulties among students when learning DP \citep{enstromdpdifficulties,shindler2022student, zehra2018student}.
We attribute this to DP's close relation to recursion and strong demand for high-level mathematical thinking.
For instance, many students who are uncomfortable with recursion find it challenging to identify the recursive patterns of the problem \citep{shindler2022student}.
Moreover, students often do not evaluate the subproblems in the correct order, resulting in redundant calculations and suboptimal worst-case runtimes \citep{enstromdpdifficulties}.

To assist students learning DP, we developed \dpvis,
a library that empowers its users with instructive and interactive visualizations of DP algorithms.
\dpvis provides three major classes of features:
(1) an instructive step-by-step visualization of the dynamic program as it finds solutions in the subproblem array, (2) an interactive self-testing feature that verifies and reinforces the user's understanding of DP, and (3) customization options for instructors to populate the visualization screen with additional annotations.

\dpvis is designed with both students and teachers in mind. 
For students, the first two types of features will likely be most useful: they allow students to use \dpvis as a convenient debugging and visualization tool, and to test their own understanding of the execution of a DP algorithm. In particular, both active engagement and leading prompts have been identified as key features towards effectiveness of algorithm visualization tools \citep{saraiya2004}.
For teachers, the third class of feature allows for easy creation of carefully annotated examples for their students' self-study and testing; again, careful instructions have been identified as a key feature of effective algorithm visualization tools.
    
The central design philosophy of \dpvis is for the programming experience to be as similar as possible to implementing a dynamic program in ``vanilla'' Python.
If a user simply wants a basic visualization of their DP array without the need for any additional features, the code is usually not substantially modified.
In many cases, the user only needs to change two lines of code to generate a visualization of the dynamic program (\autoref{fig:wis}).

We emphasize that the goal of \dpvis is not to implement visualizations for a few particular DP problems, although it provides an easy way to do so.
Instead, the library constructs a visualization of the execution of \emph{any} 1D or 2D DP algorithm, and thus also serves as an important debugging tool for anyone (in particular, students) implementing DP solutions to new problems.

We deployed \dpvis as a learning tool in an undergraduate algorithms class, and asked students about their experience in an optional survey.
Most respondents reported that \dpvis had a positive impact on their learning experience (\autoref{sec:deployment}).
We believe that the visualization features, interactive testing mode, and flexibility of \dpvis allow it to effectively assist instructors in classroom settings.

\subsection{Related Work}
\subsubsection{Existing DP Visualization Tools}

We discuss several previous DP visualization tools. Unlike other visualizers, \dpvis implements an interactive self-testing mode that quizzes the students about the outcome of the next operation performed on the subproblem array (\autoref{sec:library}).
Other visualizers \citep{BannerjiDPVisualization} may implement test cases (i.e., pass-fail checks to ensure that the implementation solves specific instances of the problem), but none place an emphasis on ensuring the student has a thorough understanding of the recurrence.

\textbf{Easy Hard DP Visualizer} \citep{EasyHardDPVisualizer} takes arbitrary 1D and 2D dynamic programs written in nearly ``vanilla'' JavaScript and shows a visualization of the subproblem array while highlighting the dependencies for each subproblem. 
However, after the initial animation, there is no option to go rewind the animation nor to pause the animation at a specific frame to view the computation in the array at any specific iteration.

\textbf{VisuAlgo} \citep{VisualGoVisualizer} illustrates dynamic programs as a directed acyclic graph (DAG) where the nodes represent the subproblems, and edges represent dependencies between them. 
VisuAlgo topologically sorts the subproblems to iteratively calculate the values of each subproblem in the correct order.
Although the graph representation clearly articulates dependencies and the ordering over the subproblems, we observed that the subproblem graph often became cluttered even for moderately sized problem instances.

Strictly speaking, both Easy Hard DP Visualizer and VisuAlgo work with recursive functions as opposed to actual dynamic programs.
They automatically perform \emph{memoization}\footnote{The process of saving the solution to subproblems is known as \emph{memoization}.} on the given recursive function and produce either an animation of an array (for Easy Hard DP Visualizer) or a DAG (for VisuAlgo).
This implicitly hides the array of subproblems, leading to ambiguity over when and where the solutions to the subproblems are stored.
By requiring that the memoization be made explicit by the user, \dpvis encourages users to have a clear understanding of their recurrence.

\textbf{Algorithm Visualizer} \citep{AlgoVisualizer} and \textbf{Dynamic Programming Visualization} \citep{BannerjiDPVisualization} offer a similar user interface as \dpvis, with a controllable frame-by-frame animation showing the subproblem array as it is populated. 
However, in Algorithm Visualizer, the programmer is expected to manually update an \emph{array tracer} with the operations that are performed on the array for each frame of the animation.
\dpvis is significantly more user-friendly by implicitly --- and non-intrusively --- tracking all \texttt{READ} and \texttt{WRITE} operations performed on the array and automatically determining which operations are visualized in which step of the animation (\autoref{sec:library}).

Dynamic Programming Visualization \citep{BannerjiDPVisualization} collects a list of DP problems and asks the students to implement a DP algorithm.
Akin to Leetcode \citep{Leetcode} and competitive programming environments \citep{Codeforces, Hackerrank}, students verify the correctness of their implementation through a set of test cases checking the final subproblem array.
Although this approach to testing verifies correctness, it does not allow students to demonstrate a thorough understanding of every step of the algorithm, and it restricts the user to the preselected set of problems.

\subsubsection{Algorithm Visualization}

Algorithm visualization (AV) is a well-studied topic in the CS education community. 
A long line of work in computer science education has provided evidence that the most effective algorithm visualizers are interactive \cite{hundhausen2002meta}.
Further explorations \cite{naps2002exploring} have provided taxonomies for the interactivity of algorithm visualizers (not viewing AVs, viewing AVs, responding to questions about AVs, changing parts of AVs, having students construct AVs, and presenting AVs). 

Studies have shown that although CS educators resoundingly agree that algorithm visualizations are effective pedagogical tools, they still hesitate to use them in classroom settings \cite{shaffer2011getting, naps2002exploring}.
Only about half of the instructors surveyed by \citet{shaffer2011getting} responded that they had used algorithm visualizers in their class over the past two years.
The two most common reasons for lack of use were ``Trouble with finding suitable AVs'' (13 responses) and ``Trouble with integrating AV material into course'' (11 responses).\footnote{For reference, the next most common reason was ``Difficulty of making AVs'' (3 responses).} 
The survey by \citet{naps2002exploring} echoes the disconnect between instructors’ beliefs and practice, with similar obstacles named.
\dpvis is designed specifically for ease of adoption to lower these hurdles.

\subsubsection{Dynamic Programming Comprehension}
Dynamic Programming comprehension has also been studied in the CS education community.
Broadly, \citet{zehra2018student} identified three types of thematic difficulties that students face when first learning DP, i.e., subproblem identification, solution technique, and defining a recurrence.
A later replication study by \citet{luu2023algorithms} replicates these findings and identifies further difficulties experienced by students. 
For instance, some of the thematic difficulties arise from a failure to comprehend the applicability of DP, an inadequate understanding of recursion, and difficulty in determining the correct order to compute subproblems.
Our library is designed to visualize computation of the subproblems, which will help students understand recursion and determine the ordering of the subproblems.

\begin{figure*}[htb]
\centering
\begin{subfigure}[t]{0.49\linewidth}
\centering
\begin{lstlisting}[language=Python,showlines=true]
arr = [-1] * n # a list of length n, initialized to -1
# sort by f[i] and precompute p[i] values (omitted here)
arr[0] = 0
for i in range(1, n):
    arr[i] = max(arr[i-1], w[i] + arr[p[i]])
        
\end{lstlisting}
\caption{Standard Python implementation without \dpvis.}
\label{fig:python_wis}
\end{subfigure}
\begin{subfigure}[t]{0.49\linewidth}
\begin{lstlisting}[language=Python,escapechar=!,numbers=none]
!\hl{arr = DPArray(n)}!
# sort by f[i] and precompute p[i] values (omitted here)
arr[0] = 0
for i in range(1, n):
    arr[i] = max(arr[i-1], w[i] + arr[p[i]])
!\hl{display(arr)}!
\end{lstlisting}
\caption{Python implementation with \dpvis visualization.}
\label{fig:dpvis_wis}
\end{subfigure}
\caption{Python implementations of the Weighted Interval Scheduling dynamic program with and without \dpvis visualizations enabled.
\dpvis visualizations can be enabled in two lines: the first line replaces the Python \texttt{List} object with the \dpvis \texttt{DPArray} object, and the second line simply generates the visualization for the \texttt{DParray} object.
Line changes from the standard Python implementation are \hl{highlighted}.}
\label{fig:wis}
\end{figure*}

\section{Background on Dynamic Programming} \label{sec:background}

Dynamic Programming (DP) is a technique taught in introductory-level algorithms courses at universities. It can be used to solve problems that otherwise do not admit straightforward polynomial-time solutions \cite{luu2023algorithms}.

The idea of DP is to find a \emph{recurrence relation} that expresses the solution to larger subproblems in terms of solutions to smaller subproblems.
By first computing and saving (``memoizing'') the solutions to the smaller subproblems, these solutions may be referenced repeatedly to solve multiple larger subproblems.
This is how DP reduces redundant calculations.

\subsection{Example: Weighted Interval Scheduling} \label{sec:wis}

We illustrate the abstract discussion of Dynamic Programming with the classic problem of finding a maximum-weight set of non-overlapping intervals, a problem we will use as a running example throughout this paper.

\begin{problem} \label{dp_ex}
    \textbf{Weighted Interval Scheduling (WIS) \cite{kleinberg2005algorithm}.}
    The input is a collection of $n$ intervals $C = \Set{(s_i, f_i, w_i)}{i \in [n]}$, characterized by their start times $s_i$, finish times $f_i > s_i$, and (positive) weights $w_i$. The goal is to find a set $S \subseteq C$ of non-overlapping intervals maximizing the sum $\sum_{i \in S} w_i$ of the weights.
    The set $S$ is non-overlapping if no two intervals in $S$ overlap, i.e., if $s_i > f_j$ or $s_j > f_i$ for each pair $i \neq j$ with $i, j \in S$.
\end{problem}

By a pre-processing sorting step, we can assume that intervals are sorted by non-decreasing finish times, i.e., $f_i \leq f_{i+1}$ for all $i$.
The subproblems under consideration will be finding the maximum WIS only for intervals $1, \dots, i$, for each $i$, i.e., acting as if only the first $i$ intervals existed.
We write $\OPT(i)$ for the maximum weight that any non-overlapping subset of intervals $1, \dots, i$ can achieve.

To cleanly express the relationship between the optimal solutions to these subproblems, let $p_i = \max\Set{j}{f_j < s_i}$, called the \emph{predecessor} of $i$, be the latest-finishing interval before interval $i$ that would not overlap with $i$.
(These $p_i$ can be pre-computed easily.)
There are two possibilities for the \Kth{i} interval: not including it or including it.
In the latter case, intervals $p_i + 1, \ldots, i-1$ are ruled out (because they overlap), but intervals $1, \ldots, p_i$ are available to be combined with interval $i$, and the optimum solution will choose an optimum subset of them.
If interval $i$ is not chosen, the optimum solution for the first $i$ intervals is the same as that for the first $i-1$ intervals. 
The optimum solution is the better of these two options.
This gives rise to the following recurrence:
\begin{align*}
    \OPT(0) & = 0 &
    \OPT(i) & = \max(\OPT(i - 1), w_i + \OPT(p_{i})).
\end{align*}

A na\"{i}ve recursive implementation would lead to exponential running time.
However, we observe that there are only $n$ total values $\OPT(i)$ that must be calculated and that $\OPT(i)$ depends only on $\OPT(i-1)$ and $\OPT(p_i)$, both smaller indices than $i$.
Thus, the optimum can be computed with the simple loop-based implementation shown in \autoref{fig:python_wis}, where the array $\texttt{arr}$ represent $\OPT$.

Once the array \texttt{arr} is fully filled in, the maximum weight is stored in \texttt{arr[n]}. Furthermore, the actual solution can be traced back in a standard way: start with index $i=n$, and see whether \texttt{arr[i] == arr[i-1]}. If so, then interval $i$ can be safely omitted in an optimal solution, and $i$ is updated to $i-1$; otherwise, $i$ must be included, which can be output, and $i$ is updated to $p_i$.

\newcommand{\READ}{%
    \definecolor{READ}{HTML}{B7609A}%
    \sethlcolor{READ}\hl{\texttt{READ}}%
    \sethlcolor{yellow}
}
\newcommand{\WRITE}{%
    \definecolor{WRITE}{HTML}{5C53A5}%
    \sethlcolor{WRITE}\hl{\texttt{WRITE}}%
    \sethlcolor{yellow}
}
\newcommand{\MAXMIN}{%
    \definecolor{MAXMIN}{HTML}{EB7F86}%
    \sethlcolor{MAXMIN}\hl{\texttt{MAX/MIN}}%
    \sethlcolor{yellow}
}

\begin{figure*}[htb]
    \centering
    \includegraphics[width=\linewidth]{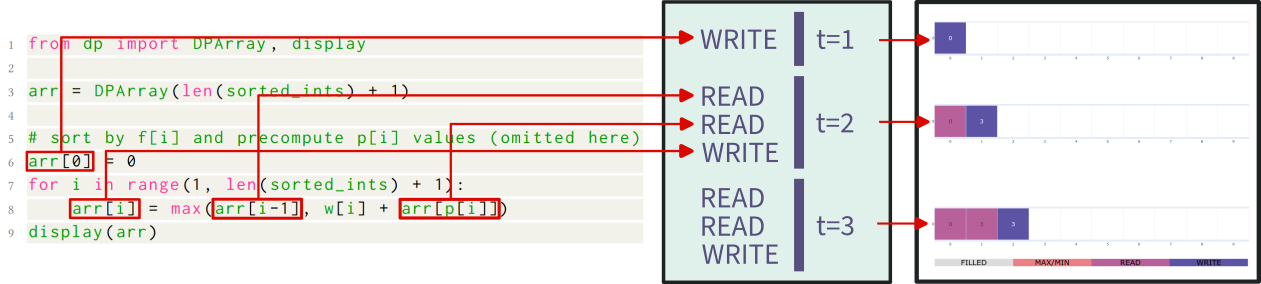}
    \caption{Each operation performed on the \texttt{DPArray} object is interpreted under one of three categories: \READ, \WRITE, and \MAXMIN (\MAXMIN is not showcased in this diagram).
    Operations are broken down into sub-sequences such that each sub-sequence ends with a \WRITE operation.
    All operations of a sub-sequence are visualized in the same frame of the animation.} \label{fig:logger}
\end{figure*}

\section{The \dpvis Library} 
\label{sec:library}

The goal of \dpvis is to help students studying dynamic programming gain a more thorough understanding, from the perspectives of both theory and implementation.
In particular, the focus on visualizing the DP array and the dependencies between its entries allows students to \emph{see} the recurrence relation in action, speeding up their debugging and learning process.

\dpvis is designed as a lightweight extension of the Python programming language, allowing a user to easily generate visualizations (including the dependencies between its cells) and self-test features for \emph{any} 1D or 2D DP array.
In particular, it serves as a convenient debugging tool for students deriving and implementing their own DP solutions.
Moreover, additional functionalities enable users to add various labels and illustrations, and thereby use \dpvis to generate pre-implemented DP algorithms for illustration and to provide to students for self-testing.

\subsection{Ease of Adoption and Use} 
The only prerequisite for using \dpvis is familiarity with the Python programming language.
We chose Python as it is beginner-friendly and a common choice in CS curricula,\footnote{\citet{onyeka2018language} finds that in 2018, 41 of 155 (26.45\%) surveyed institutions used Python in their introductory programming course for computer science students.
Moreover, Python was the most adopted language by universities that selected a new language to use in introductory courses within 2 years of the initial survey.} 
and Python is supported by easily accessible packaging systems such as Python Packaging Index (PyPI) \cite{pypi} and Anaconda \cite{anaconda}.
To make adoption as easy as possible, \dpvis has been deployed on PyPI and can be installed with a single command on any machine running Python 3.8+.

We designed \dpvis so that users can invoke its visualization functionality on a Python DP with minimal modifications to the source code.
In \autoref{fig:wis}, we invoke \dpvis visualization on a standard Python solution of the WIS problem by making \emph{two lines of change}.
First, we create a \texttt{DPArray} object in place of a Python \texttt{List} object to store solutions to prior subproblems; this automatically enables tracking of read and write operations on the array for convenient visualization.
Then, we call the \texttt{display} method on the \texttt{DPArray} object to generate the visualization.

\subsection{Main Operations of DPArray}
\dpvis tracks the operations performed on any \texttt{DPArray} during the algorithm's execution and provides a frame-by-frame animation that automatically groups relevant operations.
Currently, \dpvis supports three types of operations for \texttt{DPArray}s: \texttt{READ}, \texttt{WRITE}, and \texttt{MAX/MIN}.
A \texttt{READ} is recorded whenever an element is accessed through the bracket operator, e.g., \texttt{x = OPT[0]}.
A \texttt{WRITE} is recorded whenever a value is assigned to an index in the array, e.g., \texttt{OPT[0] = 0}.
\texttt{MAX/MIN} is a special operation that can only be performed by calling the custom \texttt{max} or \texttt{min} member functions of \texttt{DPArray}.
Their main purpose is to highlight the $\argmax$ or $\argmin$.\footnote{Recall that the $\argmax$/$\argmin$ of a function is the \textit{input} at which the function obtains its maximum/minimum value. If multiple inputs attain the max/min value, \dpvis highlights an arbitrary input.}

A major challenge in designing a visualizer for dynamic programs is determining the sub-sequence of operations to be rendered on each frame.
Algorithm Visualizer \cite{AlgoVisualizer} requires users to specify which operations are recorded in each frame in their implementation.
As a result, users must write and debug additional code that is not related to their DP algorithm.
Instead, \dpvis automatically maps operations to frames.
Given a sequence of array operations, \dpvis assumes that each \texttt{WRITE} operation signifies the end of a frame.
We reason that in most reasonable DP implementations, the algorithm only writes to the array once it has computed the solution to a subproblem, and thus the sequence of preceding operations is relevant to computing the newfound solution.
This behavior is depicted in \autoref{fig:logger} middle, where the sequence of operations is divided into three sub-sequences, each ending with a \texttt{WRITE} operation.
This design choice makes our library extremely easy to use, as users are not tasked with tracking operations themselves.

\begin{figure*}
    \centering
    \includegraphics[width=0.8\linewidth]{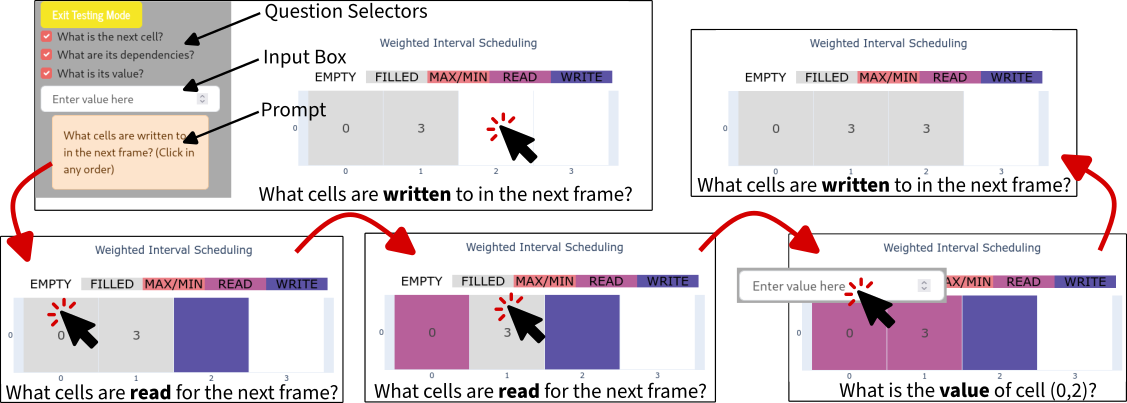}
    \caption{In testing mode, \dpvis prompts the student with a question, and the student can respond by either clicking on the array or submitting a number in the input box.
    The set of questions can be modified with the question selectors.
    Once all questions are answered correctly, \dpvis automatically advances to the next frame of the visualization and continues testing the student.}
    \label{fig:interactive_learning}
\end{figure*}

\autoref{fig:logger} illustrates a concrete example of \dpvis mapping array access and assignment operations to frames in the visualization.
In \autoref{fig:logger} left, the base case on line 6 produces a \texttt{WRITE} operation (\autoref{fig:logger} middle).
As a result, in the visualization (\autoref{fig:logger} right), the first index of the array is highlighted with the color for \texttt{WRITE} operations (i.e., dark blue).
The recurrence on line 8 (\autoref{fig:logger} left) produces the following sequence of operations: \texttt{READ} \texttt{READ} \texttt{WRITE} (\autoref{fig:logger} middle).
Since dpvis assumes that \texttt{WRITE} indicates that a solution to a subproblem has been computed, it depicts all of these operations in the same visualization frame.
Hence, for $t=2$ and $t=3$, \dpvis illustrates two \texttt{READ} operations and one \texttt{WRITE} operation.
Note that for $t=2$, the two \texttt{READ} operations are performed on the same index, so \dpvis only highlights a single cell as being read from.

As discussed at the end of \autoref{sec:wis}, it is typically straightforward to construct the actual solution (in addition to its value or cost) from the array once all entries are filled in. This process of \emph{tracing back} through the array is often very instructive in terms of case distinctions in a recurrence relation. 
\dpvis provides API functions that help visualize the traceback path; by implementing a final traceback loop with the function \texttt{add\_traceback\_path}, a user can generate a final frame showing the cells chosen in traceback; from these cells, the actual solution is easy to reconstruct.

\subsection{Interactive Learning with \dpvis}
One of the best ways to learn a topic is through spaced repetition and testing \cite{kang2016spaced}.
\dpvis automatically generates a set of questions that allows users to test their understanding on the execution of dynamic programs.

By entering ``testing mode'' in the visualization, the animation will be disabled, and the user will be prompted with questions about the state of the array in the next frame of the animation.
The user will be asked questions like ``What cells are written to in the next frame?'', ``What cells are read for the next frame?'', and ``What is the value of cell $(x,y)$?'' where $(x,y)$ is the index of the cell which is written to according to the first question (\autoref{fig:interactive_learning}).
When all questions are answered correctly, the visualization automatically advances to the next frame; then, questions for the next frame are asked.
Users may omit a subset of these questions using the question selector shown in \autoref{fig:interactive_learning}.
This feature may be useful for students to practice specific questions they struggle with.

\section{Classroom Deployment} %
\label{sec:deployment}
To preliminarily evaluate to what extent \dpvis helps students improve their understanding of DP, we deployed it in an introductory algorithms course at a major research university in the United States.
Dynamic programming is a significant part of the curriculum for this course: over nine hours of instructional time and two of ten graded homework assignments are devoted to the topic.

To supplement the learning of the 162 students taking this course, \dpvis was used in lectures multiple times to illustrate the process of filling in a DP table.
In addition, students were asked to use \dpvis to design and visualize slightly different inputs with very different optimal solutions for the well-known \textsc{Edit Distance} problem and to implement an algorithm they designed for a time allocation problem akin to \textsc{Knapsack}.
\if\arxiv1
The full text of the assignments is given in Appendix~\ref{sec:hw_questions}.
\else
(Full text of the relevant assignments is included in the arXiv version. \dkcomment{Add reference.})
\fi

\begin{figure*}
    \centering
    \includegraphics[width=\linewidth]{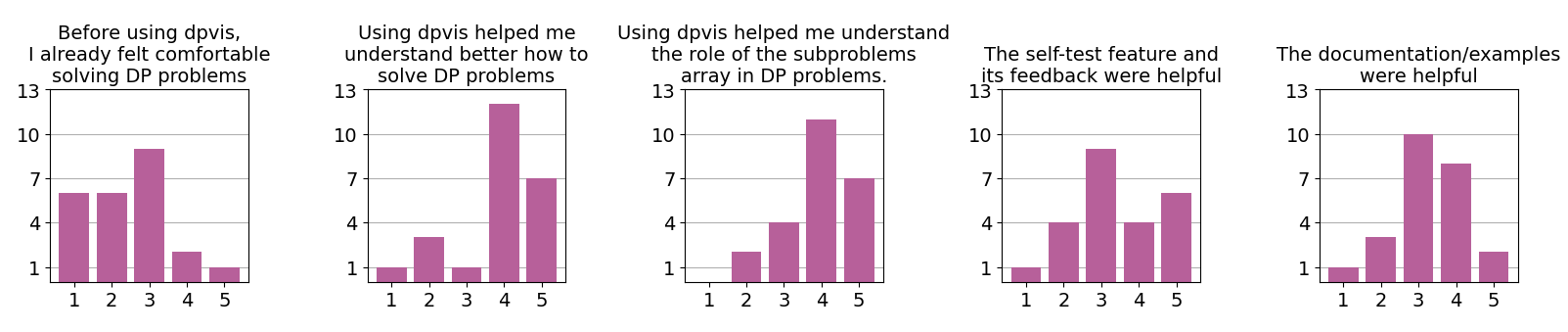}
    \caption{
        Response frequency of some questions in the survey.
        The exact prompt is displayed in the title of the plots.
        The responses are scores on a 5-point Likert scale ranging from very negative (1) to very positive (5).
    }
    \label{fig:q4-8}
    \Description{
        Response counts for questions 4--8 of the survey.
        The prompt for each question is the title of the graph. Each possible response is represented by a bar.
        The length of the bar corresponds with the number of respondents choosing the response.
    }
\end{figure*}

After the assignments utilizing \dpvis, students were asked to voluntarily provide feedback about their experience with \dpvis.
The objectives of the survey were to assess the usefulness of \dpvis as a learning tool and to identify possible improvements to \dpvis.
Students were asked to grade the usefulness of \dpvis and each primary feature (i.e., array visualization, self-testing mode, and documentation/examples) on a 5-point Likert scale from very negative to very positive.
For example, students could respond to the prompt ``The self-test feature and its feedback were helpful'' with the answers: ``very confusing'', ``somewhat confusing'', ``roughly neutral'', ``somewhat helpful'', ``very helpful.''
Furthermore, students were asked about technical difficulties (if any) they had experienced, as well as any other feedback.
A subset of the questions and their responses are shown in \autoref{fig:q4-8}.
All survey questions are provided in%
\if\arxiv1
Appendix~\ref{sec:survey_questions}.
\else
the arXiv version \dlnote{Add link}.
\fi

\subsection{Results}
\label{sec:survey_discussion}
24 students completed the survey.
\autoref{fig:q4-8} contains plots of the recorded responses to Likert scale questions;
full versions of all survey responses are included in%
\if\arxiv1
Appendix~\ref{sec:survey_results}.
\else
the arXiv version of this paper \dlnote{add link} \dkcomment{I'd add a link the first occurrence, but not the later ones.}.
\fi
In interpreting our observations, it should be noted that we received only 24 responses; therefore, the data we recorded, although suggestive, are statistically insignificant.
Second, our data were provided voluntarily, which possibly introduced self-selection bias in our samples \citep{elston2021participation}.
Lastly, since we only surveyed students from one course, cohort, and university, our data may not be reflective of all students learning dynamic programming.

\subsubsection{Difficulties Installing \dpvis}
10 respondents had technical difficulties while using \dpvis.
We believe that there are two main reasons for these difficulties.
First, Python does not have a significant role in the computer science curriculum in the university where we performed our study.
Second, at the time of the survey, \dpvis's installation process required downloading the source code, setting up a virtual Python environment \citep{anaconda}, and installing from the source code.
This involved installation process proved difficult for some students.
We have since simplified the installation, removing the need for a virtual environment to manage packages.
Now, \dpvis can be installed with a single command on any machine equipped with a compatible Python installation.
We believe that these changes make installing \dpvis easier and reduce the rate at which technical difficulties --- and with them, frustration --- occur.
In turn, the reduction in technical difficulties will hopefully lead to an increase in the number of students who find \dpvis useful.

According to \autoref{fig:q2_q5}, out of the 15 respondents who did not experience technical difficulties, 14 found \dpvis useful.
Out of the remaining eight respondents who experienced technical difficulties, five found the tool useful and three found the tool confusing.
Recent simplifications of the installation process may improve the number of students who have a positive experience with \dpvis.

\subsubsection{Effectiveness of \dpvis and its Features}
To assess whether perceived usefulness of \dpvis and its features might differ based on prior understanding of the topic, students were asked about how comfortable they felt with DP before and after experimenting with \dpvis (\autoref{fig:q4-8}).
Before using \dpvis, half of the respondents were uncomfortable with DP and nine respondents were ``somewhat comfortable''.
19 of 24 respondents found \dpvis generally useful.
The respondents identified the subproblem array visualization as the most useful feature, and 18 respondents noted that \dpvis helped them understand the role of the subproblem array.
Both the self-testing feature and documentation/examples were useful for some students.
Each feature had a positive impact on 10 respondents and no significant impact on another 9--10 respondents.

\autoref{fig:q4_q5} shows the joint frequency between how comfortable students felt with DP before using \dpvis and how much \dpvis helped their understanding.
Only a small number of respondents considered themselves comfortable or very comfortable with DP before using \dpvis, but those few had a positive assessment.
More importantly, well over 75\% of the respondents who were uncomfortable with DP before using \dpvis responded that \dpvis had made a positive impact in their studies.
Overall, it appears that \dpvis was considered mostly helpful by students with varying levels of comfort with DP before using \dpvis.

\begin{table}[t]
\centering
\setlength{\tabcolsep}{3.6pt}
\fontsize{8.1}{7.1}\selectfont
\caption{
    5-point Likert scale responses to ``I experienced technical difficulties when setting up and/or running the visualization library'' and ``Using \dpvis{} helped me understand better how to solve DP problems''.
}
\label{fig:q2_q5}
\begin{small}
\begin{tabular}{| c | c | c | c | c |} 
    \hline
        \multicolumn{2}{|c|}{
            \multirow{2}{*}{
                \begin{tabular}{@{}c@{}}Response \\ Frequency \end{tabular}
            }
        }
        &\multicolumn{3}{c|}{
            \begin{tabular}{@{}c@{}}Experienced \\ Technical Difficulties \end{tabular}
        } \\
    \cline{3-5}
        \multicolumn{2}{|l|}{}
        &\cellcolor{tableoptions} No
        &\cellcolor{tableoptions} Yes
        &\cellcolor{tableoptions} Total \\
    \hline
        \multirow{6}{*}{
            \begin{turn}{90}
            \begin{tabular}{@{}c@{}}\dpvis{} was \\ Helpful \end{tabular}
        \end{turn}}
        &\cellcolor{tableoptions} 1 & 0 & 1 & \cellcolor{tabletotal} 1 \\ 
    \cline{2-5}
        &\cellcolor{tableoptions} 2 & 1 & 2 & \cellcolor{tabletotal} 3 \\
    \cline{2-5}
        &\cellcolor{tableoptions} 3 & 0 & 1 & \cellcolor{tabletotal} 1 \\
    \cline{2-5}
        &\cellcolor{tableoptions} 4 & 8 & 4 & \cellcolor{tabletotal} 12 \\
    \cline{2-5}
        &\cellcolor{tableoptions} 5 & 6 & 1 & \cellcolor{tabletotal} 7 \\ 
    \cline{2-5}
        &\cellcolor{tableoptions} Total
        &\cellcolor{tabletotal} 15
        &\cellcolor{tabletotal} 9
        &\cellcolor{tabletotal} 24 \\
    \hline
\end{tabular}
\end{small}
\end{table}

\begin{table}[t]
\setlength{\tabcolsep}{3.6pt}
\fontsize{8.1}{7.1}\selectfont
\caption{
    Joint frequency for 5-point Likert scale responses to ``Before using \dpvis{}, I already felt comfortable solving DP problems'' and ``Using \dpvis{} helped me understand better how to solve DP problems''.
}
\label{fig:q4_q5}
\centering
\begin{small}
\begin{tabular}{| c | c | c | c | c | c | c | c |} 
    \hline
        \multicolumn{2}{|c|}{
            \multirow{2}{*}{
                \begin{tabular}{@{}c@{}}Response \\ Frequency \end{tabular}
            }
        }
        & \multicolumn{6}{c|}{
            \begin{tabular}{@{}c@{}}Comfortable with DP \\ Prior to using \dpvis \end{tabular}
        } \\
    \cline{3-8}
        \multicolumn{1}{|l}{}
        &&\cellcolor{tableoptions}1
        &\cellcolor{tableoptions}2
        &\cellcolor{tableoptions}3
        &\cellcolor{tableoptions}4
        &\cellcolor{tableoptions}5
        &\cellcolor{tableoptions}Total\\
    \hline
        \multirow{6}{*}{\begin{turn}{90}
            \begin{tabular}{@{}c@{}}\dpvis{} was \\ Helpful \end{tabular}
        \end{turn}}
        &\cellcolor{tableoptions} 1 & 1 & 0 & 0 & 0 & 0 & \cellcolor{tabletotal} 1 \\ 
    \cline{2-8}
        & \cellcolor{tableoptions} 2 & 1 & 1 & 1 & 0 & 0 & \cellcolor{tabletotal} 3 \\
    \cline{2-8}
        & \cellcolor{tableoptions} 3 & 0 & 0 & 1 & 0 & 0 & \cellcolor{tabletotal} 1 \\
    \cline{2-8}
        & \cellcolor{tableoptions} 4 & 3 & 3 & 4 & 1 & 1 & \cellcolor{tabletotal} 12 \\
    \cline{2-8}
        & \cellcolor{tableoptions} 5 & 1 & 2 & 3 & 1 & 0 & \cellcolor{tabletotal} 7 \\ 
    \cline{2-8}
        &\cellcolor{tableoptions}Total
        &\cellcolor{tabletotal} 6
        &\cellcolor{tabletotal} 6
        &\cellcolor{tabletotal} 9
        &\cellcolor{tabletotal} 2
        &\cellcolor{tabletotal} 1
        &\cellcolor{tabletotal} 24 \\
    \hline
\end{tabular}
\end{small}
\end{table}

\section{Discussion and Conclusion}\label{sec:conclusion}

In this paper, we introduce \dpvis, a Python package that seamlessly creates interactive visualizations of any DP implementation based on 1D or 2D arrays by adding/modifying as little as two lines of code (\autoref{sec:library}).
A preliminary classroom deployment suggests that \dpvis facilitates the students' understanding of DP (\autoref{sec:deployment}).

Based on our experience, we suggest some best practices for using \dpvis in an algorithms course.
These are
\begin{enumerate*}[label=(\alph*)]
    \item use \dpvis as a visualization tool in lecture, and 
    \item have students use \dpvis for a programming assignment related to DP.
\end{enumerate*}
According to our data, \dpvis was most useful for visualizing the subproblem array.
One way to leverage this feature during lectures is to visually present the execution of a DP algorithm; indeed, many lecturers and instructional videos do exactly this by hand.
Students could interact with examples provided in \dpvis when learning new DP algorithms to gain a visual intuition on how the subproblem array is computed by the recurrence relation via memoization.

Beyond visualizing \emph{given} algorithms, the visualization is especially useful for debugging DP algorithms that the students develop themselves.
To this end, \dpvis could be offered as a tool for students to debug their DP algorithms more easily than traditional, less specific, debugging tools in Python.

The most important direction for future work is a larger-scale comprehensive evaluation of the learning benefits. 
To this end, \dpvis should be deployed in more classrooms at different levels (undergraduate vs.~graduate) and at different universities. 
A higher response rate would also provide more confidence in statistical insights.
Since the installation process has been streamlined, future responses would likely not be as concerned with those roadblocks.

As with practically any education innovation, the ideal would be a randomized controlled A/B study. 
Naturally, such a study would face the typical obstacles: ethical concerns about withholding beneficial tools from some students and difficulties preventing students from using \dpvis if they hear about the benefits from their peers.

\begin{acks}
We thank Marcelo Schmitt Laser, Matthew Ferland, Gisele Ragusa, and the reviewers for helpful discussions and comments.
\end{acks}

\clearpage

\bibliographystyle{ACM-Reference-Format}
\bibliography{refs}

\if\arxiv1
\newpage
\appendix
\section{Homework Questions involving dpvis} \label{sec:hw_questions}

In this section, we provide the full text of the homework questions involving \dpvis that were assigned to students in the undergraduate algorithms class at University of Southern California.

\begin{problem}\label{hw:edit_dist}
In order to understand Dynamic Programming better, along with the tabular (for-loop) implementation and the way an optimum solution can be reconstructed, you will experiment in more depth with Edit Distance, using the dpvis library, available at 
\texttt{https://dpvis.readthedocs.io/en/latest/}.
First, download this library, and make sure you can execute one of the demos.
Notice that the implementation of Edit Distance given in dpvis has chosen a cost of 10 for insertion, 12 for deletion, and 7 for replacement. You should solve the following problem with these costs (though of course you are welcome to play with different values for your own understanding).

\begin{enumerate}[label=(\alph*)]
    \item Design three strings $x, y, y'$, each six letters long, with the following properties: (1) $y$ and $y'$ differ in exactly one replaced character, and (2) the sequence of edit operations to transform $x$ to $y$ optimally is completely different from the sequence of edit operations to transform $x$ to $y'$ optimally.
    \item Using the dpvis library, go step by step through at least one of the table calculations using self test mode, i.e., you fill in the correct values for the cells. 
    Let us know how many mistakes you made in the process, and whether you figured out how to do it correctly by the end. (There are obviously no ``wrong'' answers here.)
    \item Using screenshots, capture the final tables for the alignment of $x$ with $y$ and the alignment for $x$ with $y'$.
    Then, manually, in the tables, for each cell, mark with an arrow which of the queried cells gave the minimum cost for that cell.
    Use these arrows to recover the optimal sequences of operations (insert, delete, replace), and to illustrate how the two sequences of operations are completely different.
    Submit both the annotated screenshots and the resulting sequences.
\end{enumerate}

\begin{problem}\label{hw:time_alloc}
Imagine that you are a student at USC who has some final exams coming up in the near future.
You are currently enrolled in $n$ classes.
Your goal is to allocate your study time best between these $n$ class finals so as to maximize the GPA you get out of it.
To model this situation, you will be given the following: $H \ge 0$ is the total number of hours you have available to study; it is an integer.
For each class $i = 1, \dots, n$, the input contains an array of floating point numbers $g_i$ of size $H+1$.
For any (integer) number of hours $h$, the array entry $g_i[h]$ tells you what grade you will get in class $i$ if you study for $h$ hours.
You want to figure out how to optimally divide your $H$ study hours between the classes to maximize your GPA.
For each class, you want to study for an integer number $h_i \ge 0$ of hours, so that $\sum_i h_i \le H$.
(You do not have to exhaust your study time fully. Maybe for some classes, you get better grades by studying less.)
Your GPA with study times $h_1, h_2, \dots, h_n$ is defined as $\frac{1}{n} \cdot \sum_i g_i[h_i]$.
You will define and analyze a polynomial-time algorithm for finding the best (highest) GPA that can be achieved with any division of your study time. Your algorithm does not need to return the actual study hour assignments, only the optimal GPA.

\begin{enumerate}[label=(\alph*)]
    \item Define a suitable notion of $\OPT$ for suitably defined subproblems (with one or more parameters, as you need).
    \item Give and justify a recurrence relation for the $\OPT$ you defined.
    \item Explain what the actual final answer is and why.
    \item Turn your recurrence into an actual polynomial-time algorithm. (You do not need to prove the correctness here. We've done enough of those identical induction proofs.)
    \item Analyze the running time of your algorithm. In particular, decide if the algorithm runs in polynomial time, or just in pseudo-polynomial time. Justify your answer.
    \item Implement your algorithm using the dpvis tool, and test your algorithm to make sure that it is correct. Please include your Python dpvis code in your solution.
    \item Please provide feedback on your experience with dpvis for this and the previous homework at the following anonymous Qualtrics survey link: \textbf{Link inactive and omitted}
\end{enumerate}
\end{problem}
\end{problem}

\section{Survey} \label{sec:survey}

This section includes the full survey questions (Appendix~\ref{sec:survey_questions}) and the survey responses (Appendix~\ref{sec:survey_results}).

\subsection{Survey Questions} \label{sec:survey_questions}
\begin{enumerate}[label=\arabic*.]
    \item \emph{This question confirmed eligibility to participate in the survey.}
    \item \label{q:2} I experienced technical difficulties when setting up and/or running the visualization library.
    \begin{itemize}[label=\Square]
        \setlength{\itemindent}{5pt}
        \item Yes
        \item No
    \end{itemize}

    \item \label{q:3} If yes, what were the difficulties? \\
    \emph{This question allowed for free-response answers.}
    
    \item Before using dpvis, I already felt comfortable solving DP problems.
    \begin{itemize}[label=\Square]
        \setlength{\itemindent}{5pt}
        \item Not comfortable
        \item Somewhat uncomfortable
        \item Somewhat comfortable
        \item Comfortable
        \item Very comfortable
    \end{itemize}
    
    \item Using dpvis helped me understand better how to solve DP problems. 
    \begin{itemize}[label=\Square]
        \setlength{\itemindent}{5pt}
        \item Confused me a lot
        \item Confused me a little
        \item No change
        \item Helped me a little
        \item Helped me a lot
    \end{itemize}
    
    \item Using dpvis helped me understand the role of the OPT array in Dynamic Programming problems.
    \begin{itemize}[label=\Square]
        \setlength{\itemindent}{5pt}
        \item Confused me a lot
        \item Confused me a little
        \item No change
        \item Helped me a little
        \item Helped me a lot
    \end{itemize}
    
    \item The self-test feature and its feedback were helpful.
    \begin{itemize}[label=\Square]
        \setlength{\itemindent}{5pt}
        \item Very confusing
        \item Somewhat confusing
        \item Roughly neutral
        \item Somewhat helpful
        \item Very helpful
    \end{itemize}
    
    \item The documentation/examples were helpful.
    \begin{itemize}[label=\Square]
        \setlength{\itemindent}{5pt}
        \item N/A - didn't read
        \item Documentation was unhelpful
        \item Documentation was ok
        \item Documentation was helpful
        \item Documentation was very helpful
    \end{itemize}
    
    \item Any additional feedback, comments, or suggestions? In particular, if your experience on any of the earlier points was negative, please share details.\\
    \emph{This question allowed for free-response answers.}
\end{enumerate}

\subsection{Survey Results} \label{sec:survey_results}
All raw data of the responses are shared in\\
\url{https://drive.google.com/drive/folders/1FK0GZKRv1sKdW4MQrPtTCOIV0EMnmuwW}

\fi

\end{document}